\documentclass[proof]{WileyASNA-v1}

\articletype{Article Type}%

\received{26 April 2016}
\revised{6 June 2016}
\accepted{6 June 2016}

\newcommand{\teff}{$T_\mathrm{eff}$}

\usepackage{ulem}
\begin{document}

\title{Confirmation and characterization of neglected WDS systems using Gaia DR3 and the Virtual Observatory}
\author[1]{E. Solano}

\author[2]{I. Novalbos}

\author[3]{A. J. Ros}

\author[4,1]{M. Cortés-Contreras}

\author[1]{C. Rodrigo}

\authormark{Solano \textsc{et al}}

\address[1]{Departmento de Astrof\'{\i}sica, Centro de Astrobiolog\'{\i}a (CSIC-INTA), ESAC Campus, Camino Bajo del Castillo s/n\\
E-28692 Villanueva de la Ca\~nada, Madrid, Spain}

\address[2]{O.A.N.L. BCN, E-08032 Barcelona, Spain}

\address[3]{Asociación Astronómica de Cartagena. C/ Mayor 148, E-30394 Canteras, Cartagena, Spain}

\address[4]{Departamento de Física de la Tierra y Astrofísica. Facultad Ciencias Físicas
Universidad Complutense de Madrid. Plaza Ciencias, 1, E-28040 Madrid, Spain}

\corres{\email{esm@cab.inta-csic.es}}


\abstract{The aim of this paper is, making use of the Gaia DR3 catalogue and Virtual Observatory tools, to confirm and characterize 428 binary and multiple stellar systems classified as neglected (only one observation) in the Washington Double Star Catalogue (WDS). The components of the stellar systems have the same parallax and proper motion (within the errors) and are separated by less than 50\,000 AU, which minimizes the number of by-chance counterparts. Effective temperatures calculated using \texttt{VOSA} were used to estimate stellar masses. Binding energies were calculated for 42 binary systems confirming they are physical pairs. Also we found 75 pairs with F/G- M spectral types which are very interesting to improve the determination of the metallicity of the M star from the higher-mass component. }




\maketitle


\section{Introduction}\label{Intro}

It is well known that, while they are young, stars are not isolated but grouped in clusters, associations of stars loosely bound by mutual gravitational attraction. Cluster members share physical parameters like age and metallicity as well as kinematic properties (distances, proper motions and radial velocities). Typically, after a few hundred million years, open clusters become disrupted by close encounters with other clusters and clouds of gas, as they orbit the Galactic center. As a remnant of this process, a significant fraction of main-sequence stars (the exact percentage depends on the spectral type) are in binary and multiple systems (\citet{Duq91},\citet{Rag10},\citet{CC17}). The fact that the components of wide binary and multiple systems share physical properties and, at the same time, evolve in an independent way due to their large separation, makes them excellent testbenchs for stellar evolutionary models. In this work we aim to increase the number of these systems through examination of historical data.

The Washington Double Star Catalogue \citep[WDS,][]{Mason01} is an all-sky survey, maintained by the US Naval Observatory (USNO), that represents one of the most important databases of binary and multiple stellar systems. 
At the time of writing, the catalogue contains 154\,686 rows\footnote{\url{https://vizier.cds.unistra.fr/viz-bin/VizieR-3?-source=B/wds/wds}}. Among other information, each WDS row includes the WDS name, right ascension and declination of the primary component and the position angle and separation between the primary and secondary components. The WDS catalogue also includes a category, called {\it neglected}, to flag primaries that have been observed only once, either because the information on coordinates is wrong or simply because they have not been re-observed yet.

The Virtual Observatory (VO\footnote{\url{http://www.ivoa.net}}) is an international initiative aiming at optimizing the usage of the scientific information hosted in astronomical archives. VO has developed tools and services which enormously facilitate the access and analysis of astronomical data. In particular, Simbad \citep{Wenger00}, Vizier \citep{Ochsenbein00}, TOPCAT \citep{Taylor05}, Aladin \citep{Bonnarel00, Boch14}, and VOSA \citep{Bayo08} have been intensively used in this paper.

The paper is structured as follows. In Section\,\ref{sec2}, we describe the methodology used to obtain the sample of objects studied in this paper together with the results of their visual inspection. Physical parameters are estimated in Section\,\ref{sec3}. The Virtual Observatory SED Analyzer tool (VOSA\footnote{\url{http://svo2.cab.inta-csic.es/theory/vosa/}}) was used to compute the effective temperatures of our objects as well as to identify unresolved binaries among them. Making use of the Gaia DR3 information on colours and distances, the objects were placed on a colour - absolute magnitude diagram (CMD) allowing to separate them into main sequence objects, subgiants/giants or white dwarfs. Also, masses were estimated from effective temperatures and used to calculate binding energies. Finally, in Section\,\ref{sec4} we summarize the main results of the paper. A brief description of the Virtual Observatory compliant archive that contains
detailed information on the candidates is given in the Appendix.

\section{Sample selection}\label{sec2}

\begin{figure*}[t]
\centerline{\includegraphics[width=\textwidth, trim=0.3cm 1.5cm 0.3cm 1.5cm, clip]{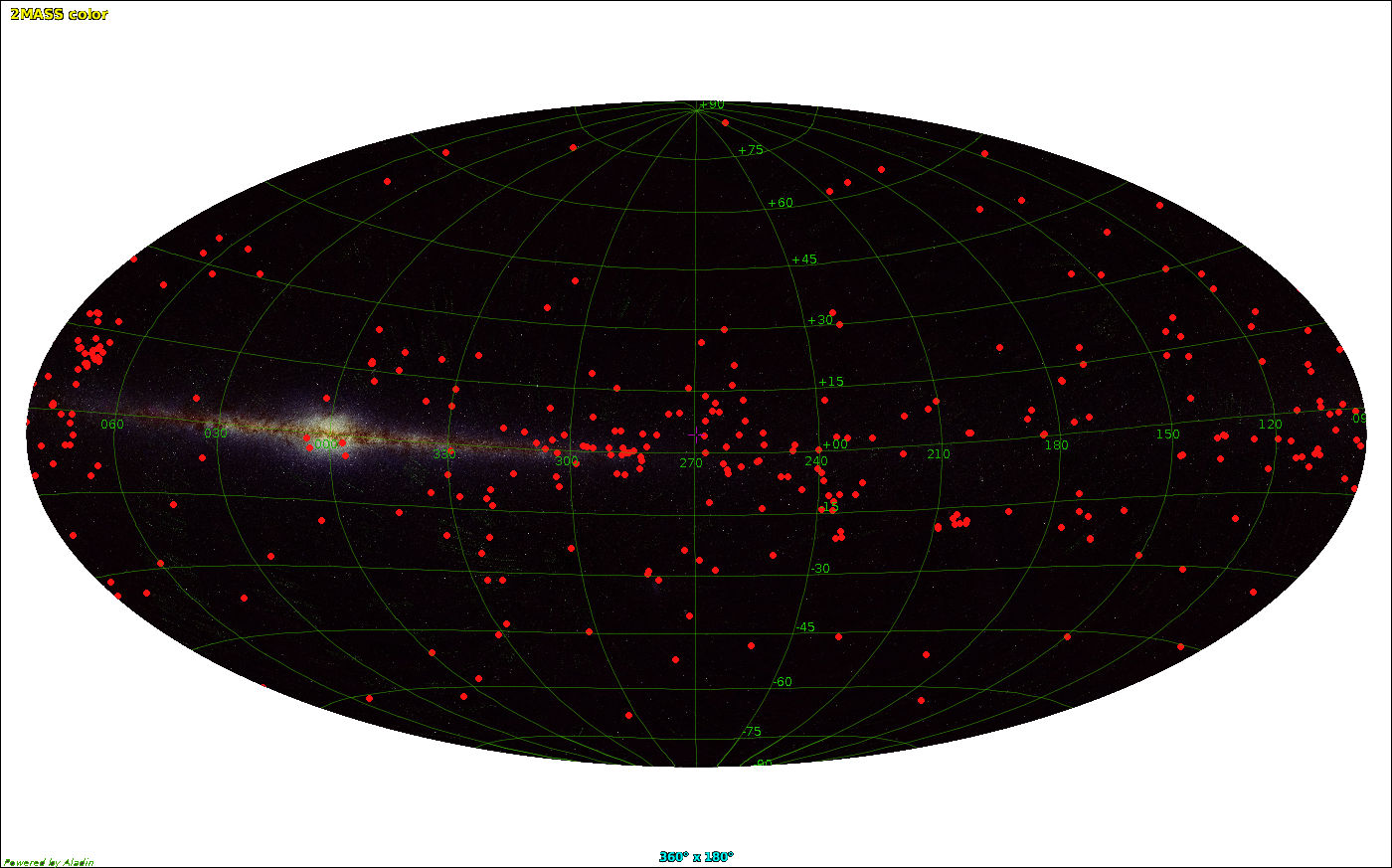}}
\caption{Sky distribution in Galactic coordinates (Aitoff projection) of the primary components of the 428 pairs identified in Section\,\ref{sec2} (filled red circles). A 2MASS coloured image is displayed in the background.\label{sky}}
\end{figure*}

\begin{figure}[t]
\centerline{\includegraphics[width=\columnwidth]{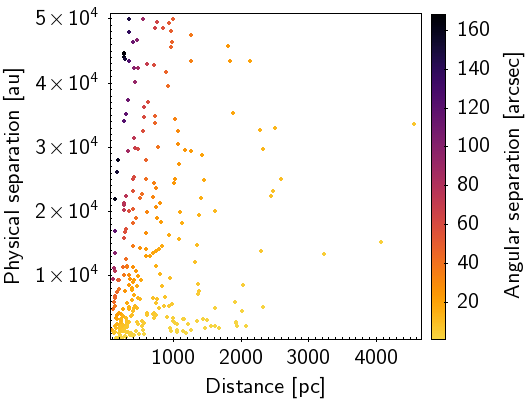}}
\caption{Separation (both physical and angular) vs distance of the 428 pairs identified in Section\,\ref{sec2}. \label{dist_sep}}
\end{figure}

The sample of objects analyzed in this paper was obtained after applying a workflow which consists of the following steps:
\begin{itemize}
 \item Filtering on the number of observations: We kept only those primaries flagged as neglected in the WDS catalogue, that is, with just one observation (N$_{obs}$=1). This search reduced the number of rows from 154\,686 to 18\,314. Some of these primaries have not been observed for many years (for instance, over two hundred were observed more than a century ago). 
 
 \item Cross-matching: The 18\,314 primary components obtained in the previous step were cross-matched with Gaia DR3. Primary components not having counterparts in Gaia at less than 5\,arcsec were rejected. If there is more than one Gaia counterpart at less than 5\,arcsec, only the nearest one was considered. The 5\,arcsec search radius was adopted as a compromise solution to avoid an unmanageable number of false positives. We also forced the association to be symmetric in the sense that the nearest object to the Gaia counterpart must coincide with our original primary component. After cross-matching the number of rows was reduced to 17\,598. 
 
 \item Filtering on parallaxes and proper motions: We used the Gaia DR3 information on parallaxes and proper motions to keep only primary components with relative errors of less than 10 per cent in $\rm{PMRA}$ and $\rm{PMDEC}$ and less than 20 per cent in parallax. The condition in parallax is necessary to have a reliable estimation of distance \citep{Luri18}. After this filtering, the number of rows reduced to 14\,808. For each one of these 14\,808 entries, we cross-matched the primary components with Gaia DR3 in a 180 arcmin radius, keeping all the counterparts that also fulfil the conditions in the relative errors in parallax and proper motion previously mentioned. We adopted this large value for the search radius to keep all the physically bound pairs. After this, we obtained 865\,974 pairs.
 
 \item Comoving pairs: From the 865\,974 pairs, we kept only those for which the differences in parallax and in proper motion, both in right ascension and declination,
 were less than 3 times the corresponding errors.  After applying this condition, 2\,735 pairs were kept. Each pair is formed by the WDS source (primary component) and a Gaia counterpart (secondary component). 
 
 \item Physical separation: According to \citet{Fran19}, the great majority of chance alignment counterparts occurs at physical separations between components larger than 50\,000 AU. Using this number as an upper limit, 603 pairs were left. Physical separations were estimated by using the formulae $s = \rho \times d$ , where $\rho$ is the angular separation between components and $d$ is the distance to the pair (estimated from parallaxes). Moreover, according to \citet{Gaiacollab21}, the minimum angular separation above which a pair can be considered by Gaia as resolved is 180\,mas. Therefore, we did not impose any condition on the minimum separation between components since this limit is much lower than what can be achieved from ground-based observations. 
 
 \item RUWE. The Gaia Renormalised Unit Weight Error (RUWE) helps to identify 
 objects with problematic astrometric solution \citep{Arenou2018, Lindegren2018, Lindegren2021}. We adopted a conservative value of RUWE $<$ 1.4 to keep stars with good astrometry. We found 429 pairs with RUWE $<$ 1.4 in both the primary and secondary components.
 
 \item Radial velocities: If a pair is  physically bound, then, the primary and secondary components should present similar radial velocities (RVs) within the errors. However, as stated in \citet{Fran19}, the radial velocity dispersion in field stars is large and the probability of chance alignment (same values within errors) is not negligible. Therefore, similar RVs cannot be used
as a proof to fully confirm that the pair is physically bound but, on the contrary, it is a good approach to discard pairs with different RV values. 

The Gaia DR3 catalog provides the RVs of both components for only 13 pairs. One of them, WDS07222-2558, showed clearly discrepant RVs (35.38$\pm$1.65 km s$^{-1}$ and 0.16$\pm$3.91 km s$^{-1}$ for the primary and secondary, respectively) and was, thus, discarded. 


\end{itemize}
After all this process, we ended up with 428 pairs comprising
 354 primaries and 372 secondaries (the same primary can be associated to more than one secondary). 
 The sky distribution of primary components is shown in Fig.~\ref{sky} while Fig.~\ref{dist_sep} provides information on the distribution of the separation between the primary and secondary components according to their distance. Detailed information on these pairs can be found at the SVO archive of neglected systems (see appendix).
 

\subsection{Visual inspection}

The 428 pairs were visually inspected taking advantage of the scripting capabilities of Aladin\footnote{\url{https://aladin.u-strasbg.fr/}}, an interactive software sky atlas. For each pair we conducted the following steps: 

\begin{itemize}
 \item Upload of an 5\,arcmin  image of the Second Palomar Sky Survey \citep[POSS II]{Reid91} centered on the position of the primary component. The POSS II image is used  as background source.
 \item Identification of the primary component using the WDS coordinates.
 
 \item Identification of the secondary component using the information on separation and position angle available in the WDS catalogue. 
 \item Upload of the Gaia DR3 sources lying in the region of the sky covered by the POSS II image.
 \item Graphical representation of the proper motion of the Gaia sources using arrows. 
\end{itemize}

\begin{figure*}[t]
\centerline{\includegraphics[width=\textwidth]{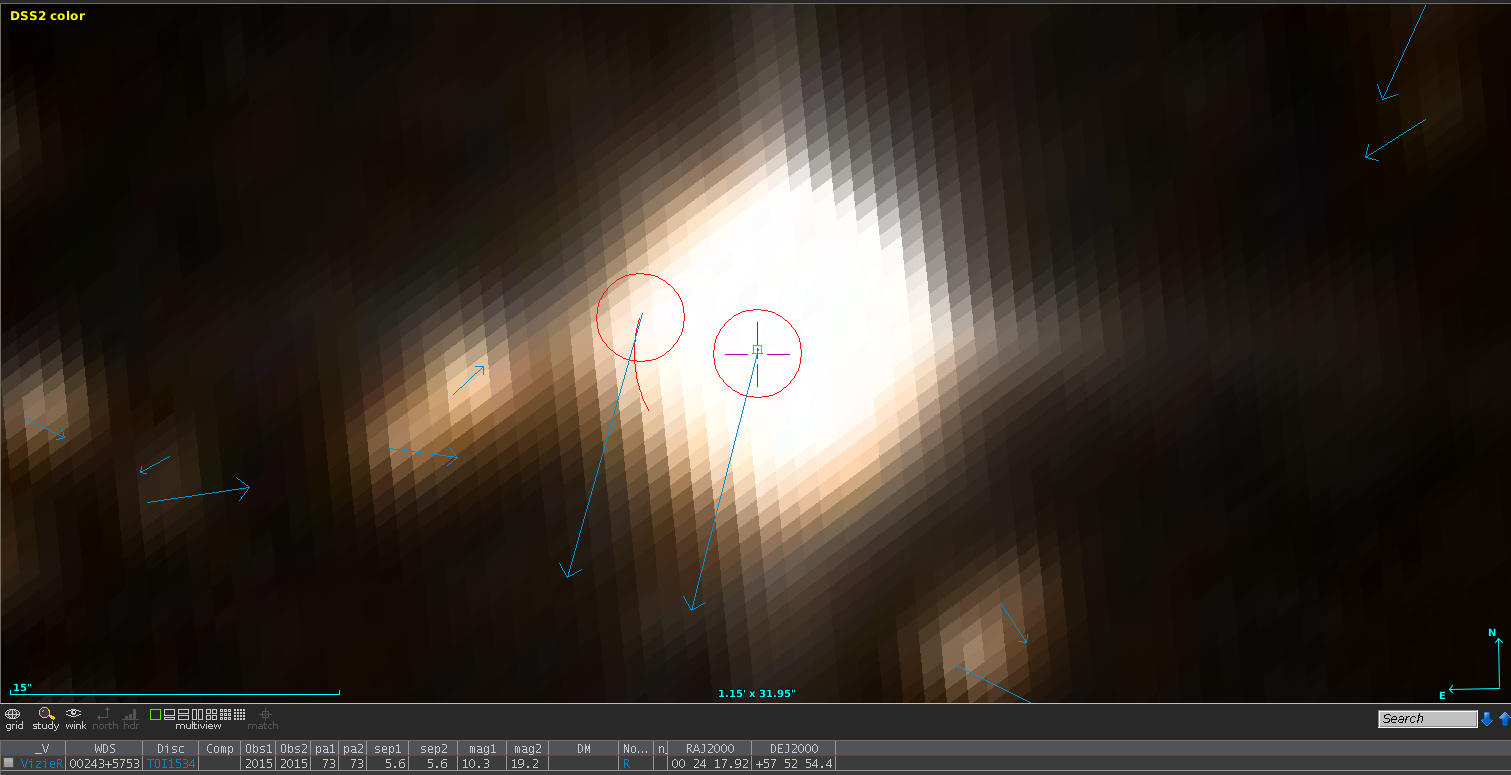}}
\caption{Example of a primary (WDS 00243+5753) for which the information provided by the WDS on separation and position angle is correct (the small red arc indicates the expected position of the secondary according to the WDS information), information that is, anyway, superseded by the Gaia parameters due to its superior performance. Blue arrows represent the Gaia DR3 proper motions. \label{fig1}}
\end{figure*}

\begin{figure*}[t]
\centerline{\includegraphics[width=\textwidth]{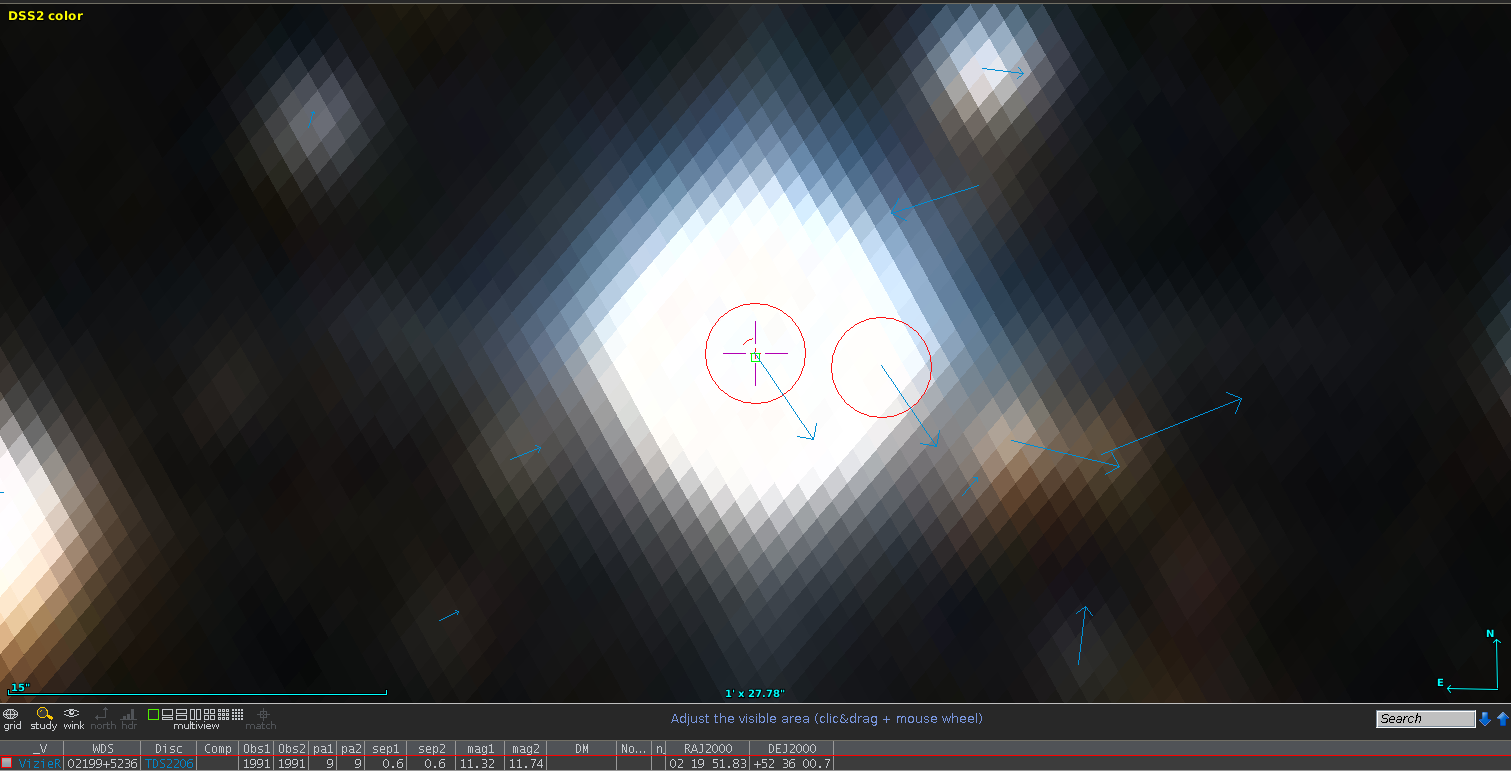}}
\caption{Example of a primary (WDS 02199+5236) for which the information provided by the WDS on separation and/or position angle is wrong. The secondary is found at a different separation/position angle (at the center of the circle that appears furthest to the right). The expected position of the secondary according to the WDS information is marked by a small red arc at the North-East of the crosshair. Blue arrows represent the Gaia DR3 proper motions.\label{fig3new}}
\end{figure*}

\begin{figure*}[t]
\centerline{\includegraphics[width=\textwidth]{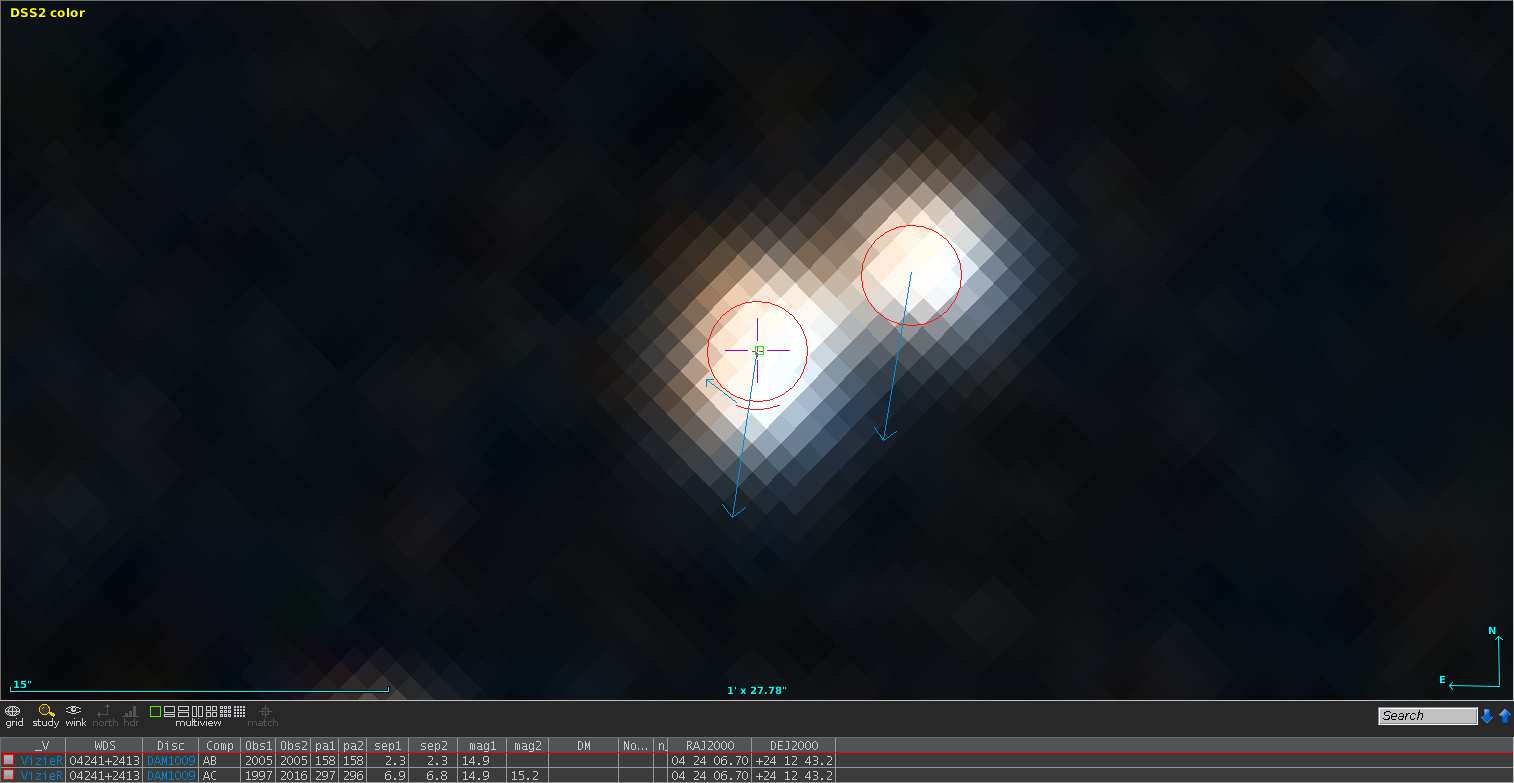}}
\caption{Example of a triple system in WDS (WDS 04241+2413) for which one of the secondaries (small red arc below the crosshair) is not physically bound based on its Gaia proper motion information (blue arrows).\label{fig4new}}
\end{figure*}

\begin{figure*}[t]
\centerline{\includegraphics[width=\textwidth]{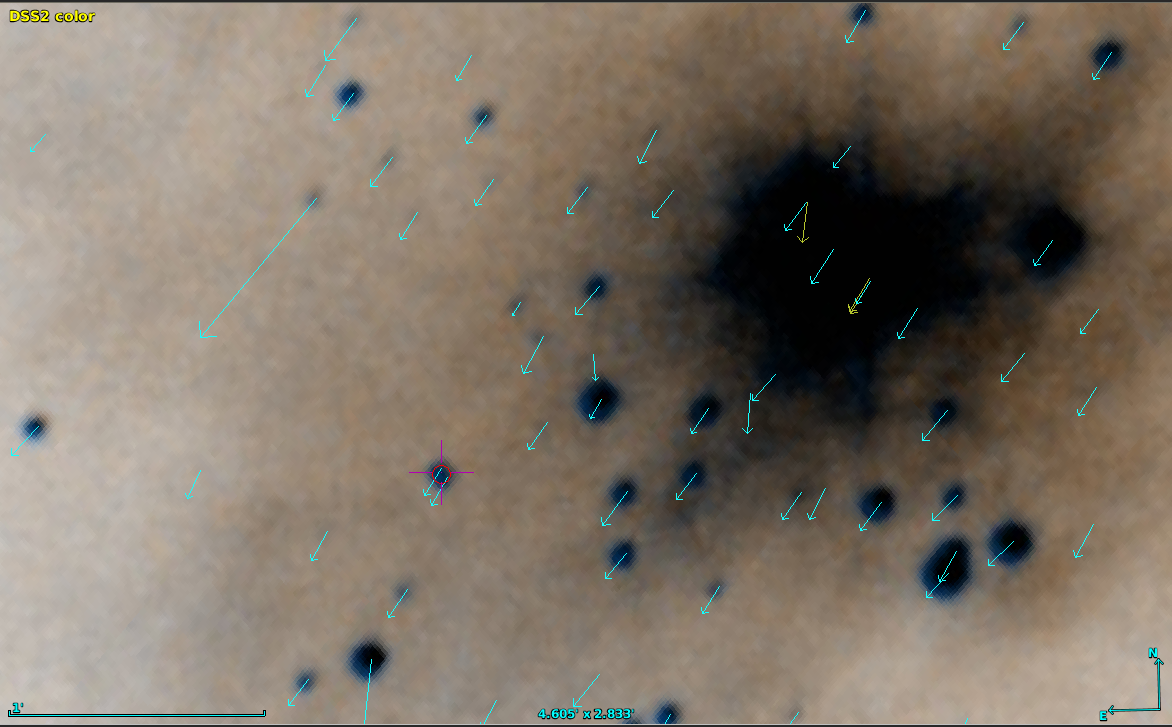}}
\caption{Example of a primary component (WDS 03447+3206, marked with a cross and a red circle) member of a stellar cluster (IC 348). \label{fig4}}
\end{figure*}

An example of the output of the script is given in Fig.~\ref{fig1}. During the visual inspection of the 428 pairs the following cases were identified:
\begin{itemize}
 \item The WDS information on coordinates of the primary component, position angle and separation is correct but it was superseded by the astrometric information provided by Gaia due to its superior accuracy (Fig.~\ref{fig1}). We found 32 pairs lying in this category.
 
 \item The WDS information is wrong. There is no Gaia DR3 source at the separation/position angle given in WDS. The secondary component is found at a different separation and/or position angle (Fig.~\ref{fig3new}). 189 pairs belong to this category. For 127 pairs, the primary component is flagged in Simbad as close binary itself. However, in all cases, they have associated a single entry in Gaia DR3 with RUWE < 1.4, indicating that they are, most likely, single. 

 \item For eight pairs there is a Gaia DR3 source at the expected position of the secondary according to the separation/position angle given in WDS. Nevertheless, the parallax/proper motion of the Gaia DR3 source lying at the expected position of the secondary are different by more than 3$\sigma$ from the parallax/proper motion of the primary component. The secondary component is, actually, found at a different separation/position angle.  

 \item 35 of our pairs belong to a multiple system according to WDS, that is, each one of the 35 primaries has associated more than one secondary. For three of the primaries we found that none of the secondaries reported in WDS have similar parallaxes and proper motions and, thus, are not physically associated. For 29 primaries only one of the secondaries reported in WDS can be considered as physical pair (Fig.~\ref{fig4new}), while we confirm that three primaries belong to a multiple system. Finally, we found three primaries, reported as double in WDS but that, actually, are triple systems according to Gaia parallaxes and proper motions. 

 \item The primary is part of a larger structure like an open cluster or a stellar association (Fig.~\ref{fig4}). 161 primaries belong to this category.
 
The previously listed categories are properly flagged in the archive (see Appendix). Also, the results obtained after the visual inspection will be reported to USNO for their ingestion in the WDS catalogue. 
\end{itemize}

\section{Physical parameters}\label{sec3}


\subsection{Effective temperatures}
 We used \texttt{VOSA} to estimate effective temperatures for our pairs. \texttt{VOSA} is a tool
developed by the Spanish Virtual Observatory designed to build
the Spectral Energy Distributions (SEDs) of thousands of objects at a time from a large number of
photometric catalogues, ranging from the ultraviolet to the infrared.
\texttt{VOSA} compares catalogue photometry with different collections
of theoretical models and determines which model best reproduces
the observed data following different statistical approaches. Physical
parameters can then be estimated for each object from the model
that best fits the data.

Using \texttt{VOSA} we queried the GALEX \citep{Bianchi17}, SDSS DR12 \citep{Alam15}, APASS DR9 \citep{Henden15}, Gaia DR3 \citep{Gaiacollab21}, 
2MASS-PSC \citep{Skrutskie06}, and ALLWISE \citep{Wright10} catalogs to build the SEDs from the ultraviolet to the infrared. Observational SEDs were then compared to the grid of BT-Settl model atmospheres \citep{Allard12}. We assumed 1\,200\,K $\leq$ $T_{\rm eff}$ $\leq$ 12\,000\,K; 3 $\leq$ logg $\leq$ 4.5 and solar metallicity. 

Extinction can play an important role in shaping the SED in particular at short wavelengths. If extinction is underestimated, the slope of the SED will appear flattened at short wavelengths and the derived effective temperature will be lower. To account for this effect and to minimize the extinction - effective temperature degeneracy in the SED fitting, we decided to leave extinction as a free parameter in the fitting process taking values in the range 0 $\leq$ Av $\leq$ 1\,mag and keep only objects at distances $<$ 1000 pc (1 Kpc roughly corresponds to an extinction of 1\,mag in the optical regime. See, for instance, Fig. 8 in \citet{Lallement22}). Therefore, effective temperatures were estimated only for this subset of objects.

The goodness of fit of the SED in \texttt{VOSA} can be assessed with the \texttt{vgfb} parameter, a pseudo-reduced $\chi^2$ internally used by \texttt{VOSA} that is calculated by forcing $\sigma(F_{\rm obs}) > 0.1\times F_{\rm obs}$, where $\sigma(F_{\rm obs})$ is the error in the observed flux ($F_{\rm obs}$). This parameter is useful to avoid the risk of overweighting photometric points with under-estimated photometric errors. Only sources with vgfb < 15 (which is an indicator of good fit) were kept.

\begin{figure}[t]
\centerline{\includegraphics[width=\columnwidth]{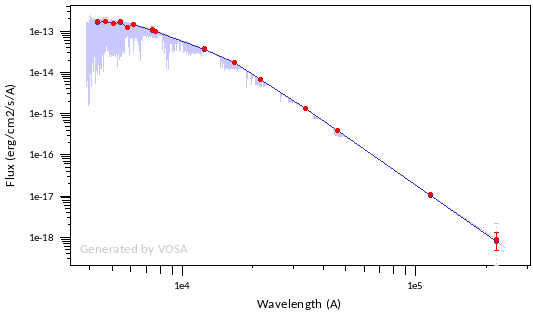}}
\centerline{\includegraphics[width=\columnwidth]{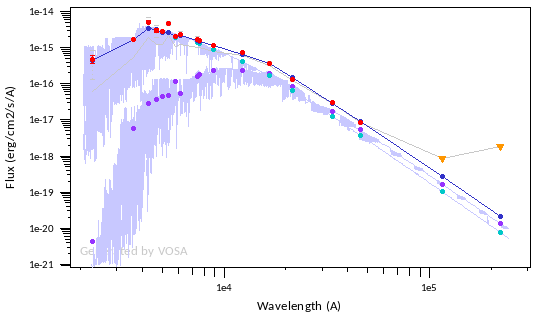}}
\caption{Top: Example of a VOSA SED fitting showing no flux excess (WDS16219-6729). The red dots represent the observed
photometry while the blue spectrum indicates the theoretical model that
fits best. Bottom: Example of a VOSA SED fitting showing flux excess (WDS06290+3717). As before, red dots represent the observed photometry, purple and green dots the synthetic photometry of the models that best fits and blue dots the composite synthetic photometry. The yellow inverted triangles indicate photometric values
corresponding to upper limits. They are not considered in the
fitting process.\label{VOSA}}
\end{figure}

VOSA also allows the identification of flux excess in a SED, excess that could be ascribed to the presence of a disc or the existence of a non-resolved companion. This way, we can check whether each of the primaries and secondaries are, themselves, single or double objects.A detailed description of how VOSA manages the flux excess can be found in the VOSA documentation\footnote{\url{https://bit.ly/2KRCv9x}}. Out of 354 primaries, VOSA did not find excess for 289 of them (216 at less than 1000 pc), while 34 primaries show flux excess, 24 of which at less than 1 Kpc. The rest of primaries (31) showed a bad SED fitting due to, for instance, poor quality photometry or lack of enough photometric points. Similarly, for the 372 secondaries, VOSA did not find excess for 256 of them (189 at less than 1000\,pc), while 43 secondaries show flux excess (32 at less than 1 Kpc). 73 secondaries were discarded because of their poor SED fitting. Physical parameters (effective temperature, luminosity, stellar radius) of the 216 primaries and 189 secondaries at less than 1000 pc and not showing flux excess can be found at the SVO archive of neglected systems (see Appendix). Examples of the VOSA SED fitting are shown in Fig.~\ref{VOSA}.

\begin{figure}[t]
		\includegraphics[width=\columnwidth]{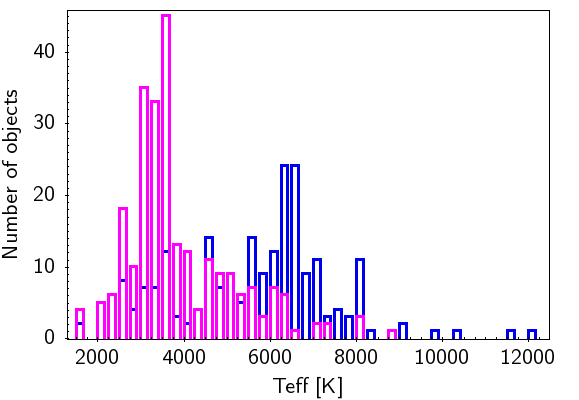}
	\caption{Distribution of the effective temperatures of the primaries (blue) and secondaries (pink) classified as single objects by VOSA.\label{teffhisto}}
\end{figure}

Fig.~\ref{teffhisto} shows the distribution in effective temperature of the primaries and secondaries classified by \texttt{VOSA} as single stars (i.e., not showing flux excess in their SEDs). The primaries reach the maximum of the distribution at $\sim$ 6\,500\,K while the secondaries have it at $\sim$ 3\,500\,K with $\sim$ 50\% of them in the range 3\,000 $\leq$ $T_{\rm eff}$ $\leq$ 4\,000\,K. Of special interest are the 75 pairs composed of a primary with F-G spectral types (5\,300\,K $<$ \teff $<$ 7\,300\,K, according to the updated version of Table~4 in \citet{pecaut13}\footnote{\url{https://www.pas.rochester.edu/~emamajek/EEM_dwarf_UBVIJHK_colors_Teff.txt}}) and an M-dwarf secondary (\teff $<$ 3\,900\,K) as the metallicity of the M-dwarf can be inferred from the hotter component.

\subsection{HR diagram}
\begin{figure}[t]
		\includegraphics[width=\columnwidth]{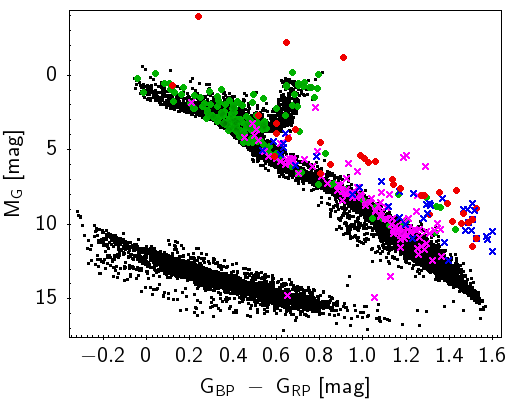}
	\caption{Colour-magnitude diagram built using Gaia DR3 sources with parallaxes larger than 10 mas, photometric errors in $G$, $G_\mathrm{BP}$ and $G_\mathrm{RP}$ less than 10 per cent and RUWE $<$ 1 (black dots). On top of it we have plotted the primary components classified by VOSA as singles (green bullets) and unresolved binaries (red bullets). Likewise, the secondary components classified as singles and unresolved binaries are plotted in pink and blue crosses, respectively \label{HRD}. The correction by the Gaia colour excess was applied following the approach described in \cite{Riello21}.}
\end{figure}

 The absolute Gaia magnitude in the $G$ band was estimated using
\begin{equation}
 M_G = G + 5 \log{\varpi} +5, 
\end{equation}
where $\varpi$ is the parallax in arcseconds and $G$ the apparent magnitude. With the absolute magnitude and the BP-RP colour, we built a colour - absolute magnitude diagram (CMD). Fig.~\ref{HRD} shows the position of our pairs in the CMD. 

Among our pairs we identified a white dwarf (Gaia DR3 150264795265471616) already reported in \citet{Fusillo19} and two sources (Gaia DR3 4009115472737321600 and Gaia DR3 6409528409064496000) lying in the locus occupied by the white dwarf - main sequence binaries \citet{alberto21}. 


\subsection{Binding energies}
\begin{figure}[t]
		\includegraphics[width=\columnwidth]{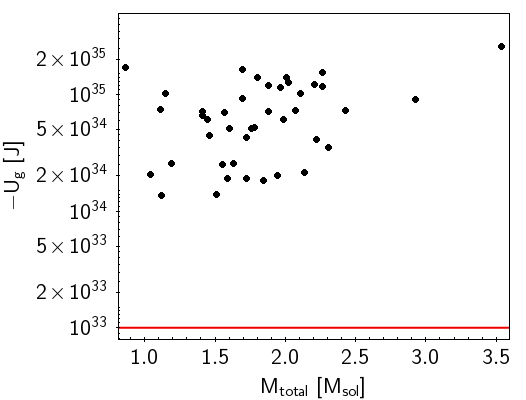}
	\caption{Binding energy-total mass diagram. The red line indicates the minimum energy above which the pair is supposed to be bound. \label{binding}}
\end{figure}

Stellar masses of objects lying on the main sequence were derived from effective temperatures by interpolating in Table~4 in \citet{pecaut13}. With these values and the projected physical separations we computed reduced binding energies as in \citet{Caballero09}
\begin{equation}
U_{g} = - G M_{1} M_{2} / s
\end{equation}
where G is the gravitational constant, M$_{1}$ and M$_{2}$ the masses of the primary and the secondary and s the projected angular separation. Fig.~\ref{binding} shows the binding enery - total mass distribution for the 42 pairs with mass determinations for both components. We consider a physically bound pair when binding energy is over 10$^{33}$\,J\,\citep{Caballero09}. All the pairs show binding energies well above this threshold.

\section{Conclusions}\label{sec4}

Starting with Washington Double Stars catalogue (154\,686 rows), we selected those primary stars having just one observation, and look for counterparts (secondaries) sharing the same Gaia parallaxes and proper motions (and radial velocities, whenever available) within the errors, and a RUWE value typical of objects with good astrometrical solution. This returned 428 pairs that were visually inspected using Aladin to check the WDS information on separation and position angle.  

Effective temperatures, luminosities and radii of both the primaries and the secondaries were estimated using VOSA. In order to minimize the impact of the effective temperature - extinction degeneracy, physical parameters were derived only for objects at less than 1\,Kpc. VOSA also allows to identify unresolved binaries by identifying the flux excess in the SED distribution. This way, out of 354 primaries, VOSA classified 289 as singles and 34 as unresolved binaries according to their SED. Similarly, for the 372 secondaries, VOSA identified 256 as singles and 43 as unresolved binaries. We were also able to identify 75 M dwarf + F/G pairs for the subsample for which the primary and secondary are both single stars. These pairs are very helpful to accurately estimate the metallicity of the M dwarf from the hotter companion. 


Using Gaia DR3 parallaxes and magnitudes, we plotted the 428 pairs on a Hertzsprung-Russell diagram (HRD). According to their position in the HRD we found a white dwarf and two sources lying in the region of the parameter space typically occupied by the white dwarf - main sequence binaries. Finally, we computed the binding energies for 42 pairs finding that all of them are consistent with being gravitationally bound. 

Detailed information of the pairs can be found in the Virtual Observatory compliant archive described in the Appendix. 

\section*{Acknowledgments}

This work has been funded by
\fundingAgency{MCIN/AEI/10.13039/501100011033} through grant \fundingNumber{PID2020-112949GB-I00} at Centro de Astrobiolog\'{i}a (CSIC-INTA). This research used the Washington Double Star Catalog maintained at the U.S. Naval Observatory. This publication makes use of VOSA, developed under the Spanish Virtual Observatory project. This research has made use of Aladin sky atlas developed at CDS, Strasbourg Observatory, France. Vizier, Simbad, and TOPCAT have also been widely used in this paper. This work has made use of data from the European Space Agency (ESA) mission {\it Gaia} (\url{https://www.cosmos.esa.int/gaia}), processed by the {\it Gaia}
Data Processing and Analysis Consortium (DPAC,
\url{https://www.cosmos.esa.int/web/gaia/dpac/consortium}). Funding for the DPAC
has been provided by national institutions, in particular the institutions
participating in the {\it Gaia} Multilateral Agreement
\appendix

\section{Virtual Observatory compliant, online catalogue}

In order to help the astronomical community with using our
catalogue of neglected WDS objects,
we developed an archive system that can be accessed from a webpage\footnote{\url{http://svocats.cab.inta-csic.es/wds_neglected/}} or through a
Virtual Observatory ConeSearch\footnote{e.g.\url{http://svocats.cab.inta-csic.es/wds_list/cs.php?RA=31.825&DEC=-7.905&SR=0.1&VERB=2}}.

The archive system implements a very simple search
interface that allows queries by coordinates and radius as
well as by other parameters of interest. The user can also select the maximum number of sources (with values from 10 to
unlimited).
The result of the query is a HTML table with all the
sources found in the archive fulfilling the search criteria. The
result can also be downloaded as a VOTable or a CSV file.
Detailed information on the output fields can be obtained
placing the mouse over the question mark located close
to the name of the column. The archive also implements the
SAMP\footnote{\url{http://www.ivoa.net/documents/SAMP}}
(Simple Application Messaging) Virtual Observatory protocol. SAMP allows Virtual Observatory applications to communicate with each other in a seamless and
transparent manner for the user. This way, the results of a
query can be easily transferred to other VO applications,
such as, for instance, Topcat.

The query syntaxes to recover the different subsets identified in this work are the following:
\begin{itemize}
    \item 32 pairs for which the WDS information has been superseded by the Gaia astrometry: \url{http://svocats.cab.inta-csic.es/wds_list/index.php?action=search} and flag=0 
    \item 189 pairs for which the WDS information on separation and/or position angle is wrong: \url{http://svocats.cab.inta-csic.es/wds_list/index.php?action=search} and flag=11 
    \item 8 pairs for which there is a Gaia DR3 source at the separation/position angle given in WDS but with a different parallax/proper motion:  \url{http://svocats.cab.inta-csic.es/wds_list/index.php?action=search} and flag=22 
    \item 3 primaries in WDS for which none of the secondaries are physical pairs according to Gaia DR3 parallaxes and proper motions: \url{http://svocats.cab.inta-csic.es/wds_list/index.php?action=search} and flag=550 
    \item 29 primaries for which only one of the WDS secondaries can be considered as a physical pair according to Gaia DR3 parallaxes and proper motions: \url{http://svocats.cab.inta-csic.es/wds_list/index.php?action=search}  and flag=551 
    \item 3 primaries, reported as part of double systems in WDS, but that actually belong to triple systems according to Gaia: \url{http://svocats.cab.inta-csic.es/wds_list/index.php?action=search}  and flag=552 
    \item Physical parameters (effective
temperature, luminosity, stellar radius) of the 216 primaries at less than 1000 pc: \url{http://svocats.cab.inta-csic.es/wds_primary}
\item Physical parameters (effective
temperature, luminosity, stellar radius) of the 189 secondaries at less than 1000 pc: \url{http://svocats.cab.inta-csic.es/wds_secondary}
\item Effective temperatures of the 75 FG + M pairs: \url{http://svocats.cab.inta-csic.es/wds_fgkm}
\item Binding energies for the 42 pairs with mass determinations for both components: \url{http://svocats.cab.inta-csic.es/wds_binding/}
\end{itemize}

\nocite{*}
\bibliography{Wiley-ASNA}%

\end{document}